# The Cooling of Coronal Plasmas. iv: Catastrophic Cooling of Loops

P. J. Cargill[1,2] and S. J. Bradshaw[3]

1. Space and Atmospheric Physics, The Blackett Laboratory, Imperial College, London SW7 2BW (p.cargill@imperial.ac.uk)

2. School of Mathematics and Statistics, University of St Andrews, St Andrews, Scotland KY16 9SS

3. Department of Physics and Astronomy, Rice University, Houston, TX 77005

Abstract We examine the radiative cooling of coronal loops and demonstrate that the recently identified catastrophic cooling (Reale and Landi, 2012) is due to the inability of a loop to sustain radiative / enthalpy cooling below a critical temperature, which can be > 1 MK in flares, 0.5 – 1 MK in active regions and 0.1 MK in long tenuous loops. Catastrophic cooling is characterised by a rapid fall in coronal temperature while the coronal density changes by a small amount. Analytic expressions for the critical temperature are derived and show good agreement with numerical results. This effect limits very considerably the lifetime of coronal plasmas below the critical temperature.

1. Introduction

The cooling phase of an impulsively heated magnetically closed coronal loop, where the heating is due to flares of any size or nanoflares, is a problem of long-standing interest. The evolution of the loop temperature and density can be used to infer long-term heating in flares (e.g. Moore et al., 1980) and, more recently, is of importance in interpreting emission measures that pertain to impulsive heating of the non-flaring active region corona (Warren et al., 2011, 2012, Schmelz and Pathak, 2012; Bradshaw et al., 2012; Reep et al., 2013).

The overall scenario is well known (e.g. Serio et al., 1991; Cargill et al., 1995). Once the peak temperature of the loop has been reached, cooling first takes place by thermal conduction with an associated evaporation of chromospheric plasma that increases the coronal density. At some point the increase in coronal density and decrease in temperature leads to cooling due to optically thin radiation becoming dominant. This radiative phase is, in fact, a combination of energy loss by radiation to space and an enthalpy flux to power the transition region (TR) radiation (Bradshaw, 2008; Bradshaw and Cargill, 2010a,b). These TR requirements lead to the loop being "over-dense" with respect to a loop in static equilibrium at the same temperature because the TR energy requirements are smaller during radiative cooling than for a static loop (e.g. Bradshaw and Cargill, 2010b; Cargill et al., 2012a). A relationship between temperature and density of the form $T \sim n^{\delta}$ holds in this radiative phase, where $\delta$ is of order 2 for hot, short loops (Serio et al., 1991; Reale et al., 1993; Cargill et al., 1995), and approaches 1 for long tenuous ones (Bradshaw and Cargill, 2010b). This scaling has been well established by one dimensional time-dependent hydrodynamic simulations.

The optically thin radiative losses take the form $n^2 R_L(T)$ ergs cm$^{-3}$ s$^{-1}$ where $n$ is the electron number density and $R_L(T)$ models the dependence of the losses on temperature. In general, either a piecewise function (Rosner et al., 1978; Klimchuk et al., 2008) or a more complete atomic physics database such as CHIANTI (Landi et al., 2012, 2013 and references therein) is used to model $R_L(T)$. In recent papers, Reale et al. (2012) and Reale and Landi (2012: hereafter RL12) incorporated a updated radiative loss function derived from the latest CHIANTI data base (Landi et al., 2012) into a 1D hydrodynamic simulation model and found faster radiative/enthalpy cooling below roughly 2 MK than seen by previous workers[1]. They also identified a final fast "catastrophic cooling" that took the loop temperature below $10^5$ K in tens of seconds. The principal cause of this behaviour was attributed to a recalculation of the physics of Fe emission lines between 1 and 2.5 MK, which lead to the coronal radiative losses in that temperature range being a factor of four larger than those of Rosner et al (1978) and perhaps twice the size of more recent loss functions (e.g. Klimchuk et al., 2008). RL12 further demonstrated that this led to much smaller emission measures in this temperature range than arise from older loss functions.

This last result is of importance in view of the relative lack of emission in the 1 – 3 MK region identified by Warren et al (2011, 2012) in some active region loops. These authors suggested that this was evidence for the active region being heated by high frequency nanoflares occurring every few hundred seconds rather than the few thousand plus seconds assumed in earlier work (Cargill, 1994). This reduced the amount of plasma cooling into the 1 – 3 MK range. RL12 proposed instead that the small amount of 1 – 3 MK plasma could be attributed to enhanced cooling, with plasma moving rapidly through this temperature range. We have used the EBTEL model (Klimchuk et al., 2008; Cargill et al., 2012a,b) to confirm this result over a range of loop conditions.

However, interpretation of the catastrophic cooling in RL12 is more subtle. While the strong losses in the CHIANTI model do lead to faster cooling below 2 MK, a final catastrophic cooling phase was also present both in our earlier simulations (Bradshaw and Cargill, 2010b) that used a single power law with modest slope to model the radiative losses ($R_L \sim T^{-1/2}$), and in the older work of Jakimiec et al (1992: Figure 1). In this paper we will explain why catastrophic cooling is generic to cooling coronal loops, and present an argument that predicts its onset as a function of loop conditions. Section 2 summarises the results of RL12 and Section 3 presents an interpretation of the "catastrophic" phase of radiative cooling.

2. The results of Reale and Landi (2012)

---

[1] It is interesting to note that Field (1965) pointed out that the onset of thermal instability is also enhanced by steeper radiative loss functions such as proposed by RL12. Of course in the present case we are not dealing with an instability, but the evolution of a system that is dynamically evolving.

RL12 ran a series of one dimensional hydrodynamic simulations of loop cooling. Figure 3 of their paper demonstrates clearly the points being discussed. The loop enters the radiative cooling phase after roughly 300 secs and cools following approximately the usual $T \sim n^2$ scaling. At 950 secs, roughly the time when the CHIANTI losses increase, the temperature suddenly falls faster, but the density adjusts to maintain approximately the $T \sim n^2$ scaling. Thus radiation/enthalpy cooling persists, and so the physical nature of the radiative cooling does not change, despite the enhanced losses. We refer to this as the intermediate cooling phase. It is only at around 0.5 MK that the $T \sim n^2$ scaling breaks down, with the loop thereafter satisfying approximately $T \sim n^4$. This final regime is referred to as catastrophic cooling. The same figure shows that a loop with the Rosner et al (1978) radiative losses obeys the $T \sim n^2$ scaling to near 0.5 MK, but without any intermediate cooling phase. Below 0.5 MK catastrophic cooling still occurs.

Figure 3 of RL12 also compares the simulation with the analytic solution of Cargill (1994), finding good agreement between the temperatures. However, the analytic solution was derived by assuming that $T \sim n^2$, so that if simulations and analytic solution are to indeed agree, the collapse in temperature found by RL12 should be paralleled by a collapse in the density. This is not what their simulations find. Taken together, these results suggest that while the updated CHIANTI losses represent an important improvement and do lead to faster cooling below 2 MK, they are not responsible for the final catastrophic cooling.

Figure 4 of RL12 shows other cases with different initial temperatures and densities as summarised in Table 1: case 3 is the example just discussed. $T_0$ and $n_0$ are the temperature and density at the start of the radiative phase, taken to be at 300 secs, and $\delta_1$ is the $T$-$n$ scaling between the start of the radiative cooling and 1.5 MK. Within the errors of reading from their figure, these values of $\delta_1$ are consistent with that expected in short loops, the loop half-length being 28 Mm in all cases.

| Case (RL12) | 1 | 2 | 3 | 4 | 5 |
| --- | --- | --- | --- | --- | --- |
| $T_0$ (MK) | 10 | 8 | 6 | 4 | 3 |
| $n_0$ ($10^9$ cm$^{-3}$) | 40 | 15 | 8 | 3.5 | 1.5 |
| $\delta_1$ | 2.3 | 2.7 | 2.2 | 2.9 | 2.5 |
| Intermediate cooling phase | Unclear | Yes | Yes | Unclear | Unclear |

| Catastrophic cooling | Yes | Yes | Yes | Unclear | No |
| --- | --- | --- | --- | --- | --- |
| $T_c$ (MK): $T^{-1/2}$ losses | 1.2 | 0.7 | 0.5 | 0.3 | 0.2 |
| $T_c$ (MK): RL12 losses | 1.4 | 1.0 | 0.7 | 0.4 | 0.2 |
| $T_c$ (MK): RL12 simulations | 1 – 1.5 | 0.5 – 0.8 | 0.4 – 0.7 | 0.2 – 0.4 | -- |

Table 1. Summary of RL12 results, as inferred from their Figures. $T_0$ and $n_0$ are the temperature and density at the start of the radiative phase (roughly 300 secs), $\delta_1$ is the slope of the $T$-$n$ relation between the start and 1.5 MK. The following two rows comment on the existence of an intermediate cooling phase starting around 1.5 MK, and a catastrophic cooling phase. The final three rows are the critical temperatures ($T_c$) for the onset of catastrophic cooling based on a radiative loss scaling as $T^{-1/2}$ (Bradshaw and Cargill, 2010b), a parameterisation of the losses shown in Figure 1 of RL12, and the range of $T_c$ inferred from Figure 4 of RL12. These last three rows are discussed in Section 3.

Case 1 has the highest density and shows no clear evidence for the intermediate phase and instead moves straight to the catastrophic regime. Case 2 is similar to case 3. Cases 4 and 5 show no clear evidence for a change to the intermediate cooling phase near 1.5 MK, but case 4 might be undergoing catastrophic cooling onset near 0.5 MK. Case 5 shows no evidence of catastrophic cooling, as also suggested by RL12 Eq (10). The question now arises whether there is any unifying physics that can account for these results.

3. Results

Bradshaw and Cargill (2010b: BC10) used a one-dimensional hydrodynamic code that models the plasma properties along a magnetic flux element to study the radiative cooling of loops with a wide range of initial temperatures, densities and loop lengths, as documented in Table 1 of that paper. They used a single power law radiative loss function: $R_L(T) = 2.19 \; 10^{-19} \; T^{1/2}$ ergs cm$^3$ s$^{-1}$ above, and $1.09 \times 10^{-31} \; T^2$ ergs cm$^3$ s$^{-1}$ below $10^{4.93}$ K, so that the increase in the losses present in RL12 below 2 MK was not present. The radiative loss at $10^6$ K is roughly half of that used by RL12. The loop temperature and density structure prior to the cooling phase was created by imposing a constant heating function for several thousand seconds, allowing the coronal temperature and density to adjust to an equilibrium. The cooling was then initiated by turning this heating off. Further details can be found in BC10.

The four panels of Figure 1 in this paper reproduce the centre-left panels of Figures 1 – 4 in BC10, showing the relationship between the average T and average n for four groups of five loops (the averages are over the middle 50% of the entire loop and apex quantities give very similar results). Time increases as each curve is followed in the clockwise direction. The important difference between the panels is the loop length, defined as being the distance between the two chromospheres. Defining L as the loop half-length, panels (a) – (d) have 2L in the range 30 – 35, 67 – 72, 105 – 110 and 205 – 210 Mm respectively. The small differences arise because the loops in each group of five were chosen to have the same length prior to the heating being turned on, and hotter, denser loops force the top of the chromosphere further down. So, within each sub-group, the shortest (longest) loop has the lowest (highest) temperature and density prior to cooling. Further details are in BC10.

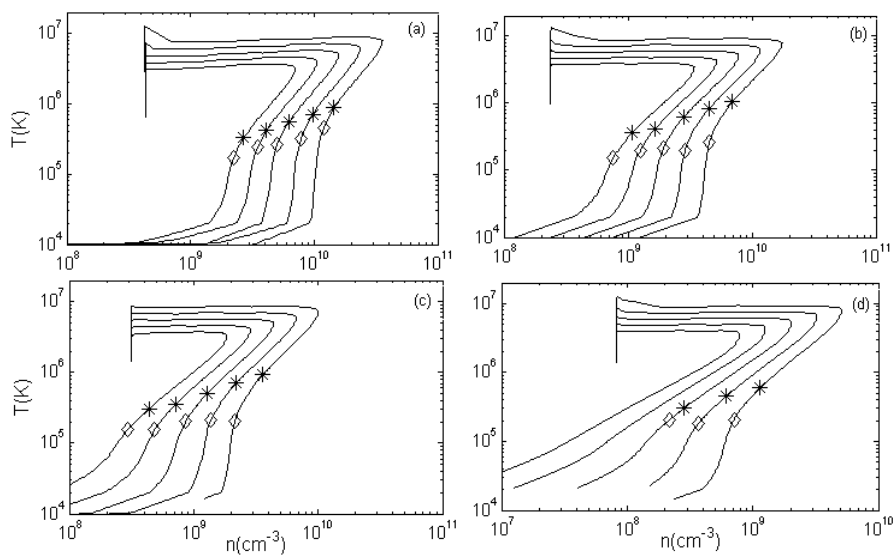

Figure 1. The relation between the average temperature and density for the 20 loops discussed in BC10. Time increases when each curve is followed in a clockwise direction. The four panels, labelled (a) – (d), correspond to groups of five loops with lengths (2L) in the range 30 – 35, 67 – 72, 105 – 110 and 205 – 210 Mm respectively (see text for details). Within each grouping, the case numbers from BC10 increase from right to left (i.e. starting on the right of panel (a), cases 1 – 5 are shown). The lower numbered cases in each panel start the radiative cooling with higher temperature and density. The stars and diamonds show the start and end of the transition to catastrophic cooling. In panel (d), the two leftmost cases do not undergo catastrophic cooling.

Figure 1 shows the following well known features. The horizontal lines correspond to the loop attaining static equilibrium, with the density increasing due to evaporation of chromospheric material. When the heating is turned off, the temperature and density both fall, and the loop enters its enthalpy/radiation cooling phase, with a range of values of the coefficient δ as documented in Table 2 of BC10. However, there is a temperature between 1 MK and 0.1 MK, depending on the case, where this *T-n* scaling starts to break down. The temperature at which this happens is the onset of catastrophic cooling and is shown in Figure 1 by a star, except for the two cases where catastrophic cooling does not occur. At a lower temperature, shown by the diamond symbol, the transition to catastrophic cooling is complete, *δ* becomes larger, as was also found by RL12, and the loop now cools with relatively little change in the density. Above the starred temperature, the enthalpy flux to the TR regulates the rate of coronal cooling by decreasing the density. Below this temperature, the diminution of the enthalpy flux gradually leads to faster coronal cooling, both due to the higher coronal density and the increasing radiative losses as the temperature falls. So the physics behind the "catastrophic cooling" of RL12 involves both the new higher losses and the diminution of the enthalpy flux from the loop. Note also that after the temperature reaches around 20,000 K, there is a rapid draining of the loop, since it is no longer possible to sustain the high coronal density in hydrostatic equilibrium.

However, as in RL12, there are cases in panels (c) and (d) where the transition to catastrophic cooling is not obvious, and the radiative/enthalpy cooling persists to lower temperatures, indeed the cases in panel (d) with the lowest initial temperatures show no evidence of this transition at all: these are long loops with low initial temperature and density. Thus, there is a dependence on the onset of catastrophic cooling on the loop length as well as the initial temperature and density of the loop at the start of the cooling.

To understand these results, we note that in the radiation/enthalpy phase the need to power the TR radiation by an enthalpy flux from the corona leads to a weak deviation from hydrostatic equilibrium that sets up the required downflow. The magnitude of the downflow continually adjusts through sound waves, permitting the corona to drain while maintaining the relationship between *T* and *n*. This regime holds provided the radiative cooling time in the corona ($\tau_R$) is longer than the sound travel time ($\tau_S$):

$$\tau_R = \frac{3k_B T}{n R_L(T)} > \tau_s = \frac{L}{C_s}, \tag{1}$$

where for simplicity averaged quantities are assumed. Here $C_s = (2k_B T/m_p)^{1/2}$ is the isothermal sound speed for an electron-proton plasma with $k_B$ and $m_p$ being Boltzmann's constant and the proton mass respectively. When Eq (1) is violated, communication between TR and corona is interrupted, the radiation/enthalpy cooling phase will terminate, the loop cools predominately by radiation, eventually leading to the catastrophic cooling seen in the simulations. The condition for the onset of catastrophic cooling is then:

$$\frac{\tau_s}{\tau_R} = \left(\frac{\tau_{s0}}{\tau_{R0}}\right)\left(\frac{T_0}{T}\right)^{3/2}\left(\frac{n}{n_0}\right)\left(\frac{R_L}{R_{L0}}\right) > 1 \tag{2}$$

where subscript "0" denotes the start of the radiative cooling phase.

If we now assume the scaling $T \sim n^2$, (2) can be written as:

$$\left(\frac{T}{T_0}\right)\left(\frac{R_{L0}}{R_L}\right) < \frac{\tau_{s0}}{\tau_{R0}} \tag{3}$$

Since the temperature falls and the radiative losses rise during cooling, (3) suggests that <u>every</u> cooling coronal loop with a sensible loss function could try to enter such a catastrophic cooling phase. A sudden rapid increase in $R_L$ as proposed by RL12 can enhance the onset. For a single power-law loss function $R_L(T) = \chi T^\alpha$, (3) becomes:

$$\left(\frac{T}{T_0}\right)^{1-\alpha} < \frac{\tau_{s0}}{\tau_{R0}} \tag{4}$$

where $\frac{\tau_{s0}}{\tau_{R0}} = \frac{L n_0 \chi}{3 k_B T_0^{3/2-\alpha} \sqrt{2k_B/m_p}} = 1.88 \; 10^{11} \chi \frac{L n_0}{T_0^{3/2-\alpha}}$ This gives a critical temperature for the commencement of the transition to catastrophic cooling as $T_c = T_0 \left(\frac{\tau_{s0}}{\tau_{R0}}\right)^{1/(1-\alpha)}$. For more general cooling with $T \sim n^\delta$, (4) is:

$$\left(\frac{T}{T_0}\right)^{3/2-1/\delta-\alpha} < \frac{\tau_{s0}}{\tau_{R0}} \tag{5}$$

A test of this comes from Figure 1. We have used the loop half-lengths, initial values of $T$ and $n$ (Table 1 of BC10) and value of $\delta$ in the radiative cooling phase (Table 2 of BC10) to calculate the temperature at which the transition to catastrophic cooling starts using Eq (5), and compared this with the estimates from the simulation results of the temperature at which the cooling (a) starts to deviate from the radiation/enthalpy phase and (b) enters the full catastrophic cooling phase, as indicated by the stars and diamonds respectively in Figure 1. The former is the relevant comparison since it is at that time that the sound waves begin to lose the ability to sustain the enthalpy flux from the corona. Figure 2 shows the results: the critical temperatures from Eq (5) are circles, and the stars and diamonds are the same as in Figure 1. The horizontal axis shows the case number from BC10.

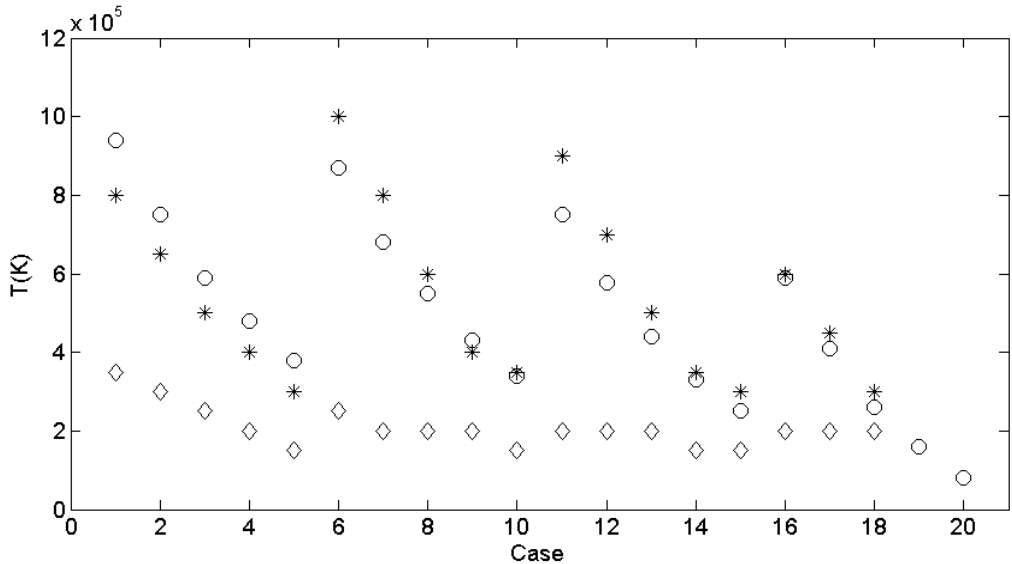

Figure 2 Temperatures associated with the breakdown of radiation/enthalpy cooling for the 20 cases discussed in BC10. Circles are analytic estimates using Eq (5), stars and diamonds are taken from the simulations shown in Figure 1, and denote the start and end of the transition to catastrophic cooling. Cases 19 and 20 show no evidence for catastrophic cooling in the numerical results.

The analytic estimates for breakdown of the radiation/enthalpy phase by-and-large agree well with the numerical estimates. It should also be noted that the trend of this critical temperature as the loop parameters change is also reproduced. Note that the very long loops do not really ever enter a catastrophic cooling phase: while the sound travel time is long, the radiative cooling time is even longer due to the low density.

This model can be applied to the results of RL12, as documented in the last three rows of Table 1, though a single power-law radiative loss function cannot in general be used. However, we can model the RL12 losses by using a loss function $R_L = 4\ 10^{-10}\ T^{-2}$ ergs cm$^3$ s$^{-1}$ between 3 MK and 1 MK and 4 x $10^{-22}$ ergs cm$^3$ s$^{-1}$ below 1 MK and solving Eq (5) iteratively. In applying Eq (5) we take L = 28 Mm which is the RL12 half-length minus their chromosphere. The three rows show the critical temperature at which the transition to catastrophic cooling begins using (a) our loss function, (b) the above approximation to the enhanced CHIANTI one, and (c) an estimate of the range of $T_c$ from the simulations of RL12. The agreement between simulations and analytic model is again satisfactory, showing the correct trends as $T_0$ and $n_0$ vary.

4. Conclusions

This paper has demonstrated that, while the temperature evolution in a cooling loop is indeed changed by upgraded (and enhanced) radiative losses such as those in CHIANTI proposed by RL12, the basic physics of radiation/enthalpy cooling is not changed in the temperature regime where the changes are most significant. Instead, we believe that the work of RL12 and BC10 shows the onset of a transition to a catastrophic cooling typically below 1 MK, though sometimes higher for hot loops with the CHIANTI losses, at a temperature determined by the inability of sound waves propagating within the loop to sustain the radiation/enthalpy cooling. This onset occurs at higher temperatures for short, hot loops such as might arise in compact flares, at temperatures between 1 MK and 0.5 MK for loops such as seen in active region cores, and at temperatures approaching 0.1 MK for long tenuous loops. This implies that the total loop cooling time from peak temperature to chromospheric values is decreased below standard values (e.g. Cargill et al., 1995) for flares and ARs, but is unchanged for long and high loops.

Other consequences are as follows. As discussed by RL12 the new CHIANTI losses do indeed lead to a lower emission measure in the region 1 – 3 MK with consequences discussed earlier. However, as we show elsewhere (Cargill, 2013), the change in the slope is not adequate to account for the range of *EM-T* profiles seen, though further atomic physics uncertainties (Bradshaw et al., 2012; Reep et al., 2013 Guennou et al., 2013) may weaken that conclusion. Secondly, catastrophic cooling below 1 MK will reduce very significantly any emission from those temperatures, at least from the coronal portion of cooling loops. This discounts further the option that the corona can account for this awkward region of the emission measure, which shows a strong upturn below 0.5 MK (see Klimchuk, 2012 for a further discussion of this problem).

To conclude, in this latest paper of our series we can state with some confidence that there now seems to be a fairly complete picture of how coronal loops cool (see also Serio et al., 1991; Jackimiec et al., 1992; Reale et al., 1993; Cargill et al., 1995; Bradshaw and Cargill, 2006, 2010a,b; Reale and Landi, 2012). Following termination of heating there are four phases: (1) Conductive cooling and associated density increase until (2) the density is large enough for radiation/enthalpy cooling to take over until (3) the enthalpy cooling is supressed and the loop cools rapidly by radiation at roughly constant density and (4) a final draining of the highly over-dense loop. Important changes in the coronal radiative losses as in RL12 can change the temperature at which these various transitions occur, but not the overall sequence.

Acknowledgements

SJB thanks the NASA Supporting Research and Technology Program. We are grateful to the International Space Science Institute (ISSI) for supporting our team, to Helen Mason for acting as co-leader of this team with SJB, and to Fabio Reale for discussions of his results at our team meetings and comments on the manuscript.